\documentclass[manuscript,screen,sigconf
]{acmart}
\AtBeginDocument{%
  \providecommand\BibTeX{{%
    \normalfont B\kern-0.5em{\scshape i\kern-0.25em b}\kern-0.8em\TeX}}}

\copyrightyear{2021} 
\acmYear{2021} 
\setcopyright{acmcopyright}\acmConference[ESEM '21]{ACM / IEEE International Symposium on Empirical Software Engineering and Measurement (ESEM)}{October 11--15, 2021}{Bari, Italy}
\acmBooktitle{ACM / IEEE International Symposium on Empirical Software Engineering and Measurement (ESEM) (ESEM '21), October 11--15, 2021, Bari, Italy}
\acmPrice{15.00}
\acmDOI{10.1145/3475716.3484187}
\acmISBN{978-1-4503-8665-4/21/10}

\usepackage{booktabs}
\usepackage{makecell}

\begin{document}

\title{
%
%
Web Application Testing: Using Tree Kernels to Detect Near-duplicate States in Automated Model Inference
%
%
}

\author{Anna Corazza}
\authornotemark[1]
\email{anna.corazza@unina.it}
\orcid{0000-0002-9156-5079}
\affiliation{%
  \institution{Università degli Studi di Napoli Federico II}
  \streetaddress{Via Claudio, 21}
  \city{Naples}
  \country{Italy}
  \postcode{80125}
}

\author{Sergio Di Martino}
\authornotemark[1]
\email{sergio.dimartino@unina.it}
\orcid{0000-0002-1019-9004}
\affiliation{%
  \institution{Università degli Studi di Napoli Federico II}
  \streetaddress{Via Claudio, 21}
  \city{Naples}
  \country{Italy}
  \postcode{80125}
}

\author{Adriano Peron}
\authornotemark[1]
\email{adrperon@unina.it}
\orcid{0000-0002-7111-3171}
\affiliation{%
  \institution{Università degli Studi di Napoli Federico II}
  \streetaddress{Via Claudio, 21}
  \city{Naples}
  \country{Italy}
  \postcode{80125}
}

\author{Luigi Libero Lucio Starace}
\authornote{All authors contributed equally to this research.}
\email{luigiliberolucio.starace@unina.it}
\orcid{0000-0001-7945-9014}
\affiliation{%
  \institution{Università degli Studi di Napoli Federico II}
  \streetaddress{Via Claudio, 21}
  \city{Naples}
  \country{Italy}
  \postcode{80125}
}

\begin{abstract}

\textbf{Background:} In the context of End-to-End testing of web applications, automated exploration techniques (a.k.a. crawling) are widely used to infer state-based models of the site under test.
These models, in which states represent features of the web application and transitions represent reachability relationships, can be used for several model-based testing tasks, such as test case generation. 
However, current exploration techniques often lead to models containing many near-duplicate states, i.e., states representing slightly different pages that are in fact instances of the same feature.
This has a negative impact on the subsequent model-based testing tasks, adversely affecting, for example, size, running time, and achieved coverage of generated test suites.
\textbf{Aims:}
As a web page can be naturally represented by its tree-structured DOM representation, we propose a novel near-duplicate detection technique to improve the model inference of web applications, based on Tree Kernel (TK) functions. TKs are a class of functions that compute similarity between tree-structured objects, largely investigated and successfully applied in the Natural Language Processing domain.
\textbf{Method:}
To evaluate the capability of the proposed approach in detecting near-duplicate web pages, we conducted preliminary classification experiments on a freely-available massive dataset of about 100k manually annotated web page pairs. We compared the classification performance of the proposed approach with other state-of-the-art near-duplicate detection techniques.
\textbf{Results:}
Preliminary results show that our approach performs better than state-of-the-art techniques in the near-duplicate detection classification task.
\textbf{Conclusions:}
These promising results show that TKs can be applied to near-duplicate detection in the context of web application model inference, and motivate further research in this direction to assess the impact of the technique on the quality of the inferred models and on the subsequent application of model-based testing techniques.

\end{abstract}




\maketitle

\section{Introduction}\label{sec:intro}

Web applications have become pervasive and are involved in many aspects of our daily lives.
From home banking to public transit trip planning, from e-commerce to social networks, society relies on web applications to an ever-growing extent for a multitude of economic, social, and recreational activities.
The impact of failures in a web application may range from simple inconveniences for end-users up to complete business interruption, and can potentially cause significant damages.
Hence, ensuring the quality and correctness of web applications is of undeniable importance \cite{ricca2019three}.

End-to-end (E2E) web testing is one of the main approaches to ensure the quality of web applications. In this kind of activity, testers exercise the Application Under Test (AUT) as a whole, from the perspective of an end-user interacting with the Graphical User Interface (GUI), i.e., the web pages, of the application. 
The goal is to verify that the web application behaves as intended in response to user-generated events and interactions with the GUI (e.g., clicks, scrolls, forms filling and submissions, etc.).
To do so, testers typically develop test scripts that, leveraging test automation libraries such as Selenium \cite{bruns2009web}, automate the set of manual operations that the end-user would perform on the GUI of the web application.
Developing such test scripts manually, however, is a time consuming and expensive task, often neglected in web projects because of resource constraints \cite{chaini2015test}. 
To support these E2E web testing activities, several model-based approaches have been proposed in the software engineering community, including test case generation \cite{biagiola2019diversity,biagiola2017search,marchetto2008state,andrews2005testing,ricca2001analysis} and test artifact generation \cite{stocco2017apogen,stocco2016clustering}. 
To infer the AUT models underlying these approaches, automated exploration techniques \cite{yandrapally2020near}, also referred to as web application crawling \cite{leithner2020xiev}, are widely used. Broadly speaking, these crawling-based techniques dynamically and systematically analyze the AUT starting at an initial page, and then explore the application by generating GUI events and checking the responses. 
When, as a consequence of a fired event, changes in the web page are detected, a new state is added to the model. In these models, a state represents a web page of the application, and transitions between states represent the fact that the target state is reachable from the source one under particular conditions (e.g., when a particular event is fired).

\begin{figure}
    \centering
    \includegraphics[width=.9\linewidth]{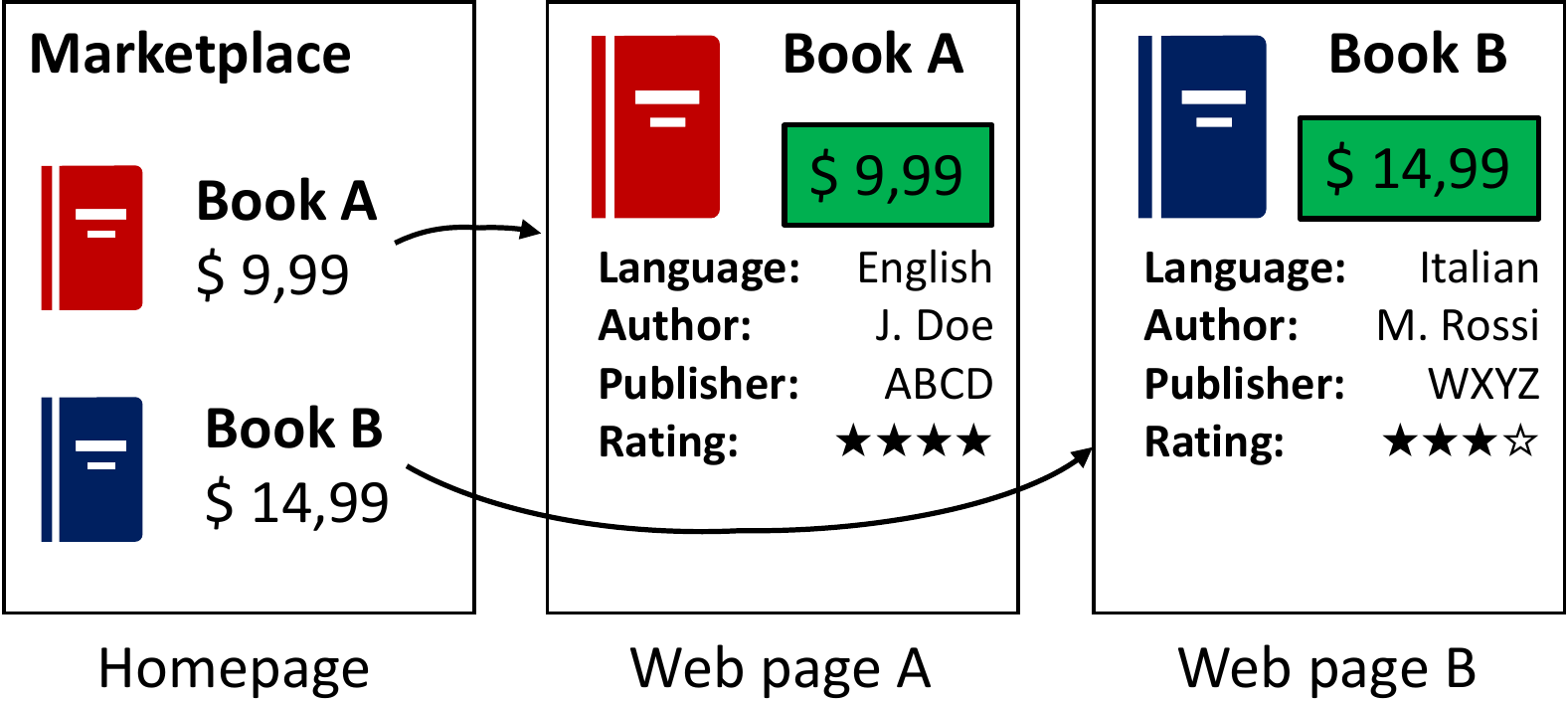}
    \caption{Example of near-duplicate web pages}
    \label{fig:near-duplicates}
\end{figure}

From a testing view-point, these inferred models should contain a minimal set of significantly different states, yet adequate to cover all the functionalities of the AUT. 
In practice, however, models inferred automatically through crawling are often affected by \textit{near-duplicates} \cite{fetterly2003evolution,dilucca2001clone,henzinger2006finding,manku2007detecting}, i.e., replicas of the same functional web page differing only by minor changes \cite{yandrapally2020near}.
As an example, let us consider Figure \ref{fig:near-duplicates}, in which three web pages from an imaginary bookstore web application are depicted. 
The homepage of the application shows a catalog of available books. After clicking on one of the books, the user is redirected to a detail web page with additional information, from which it is possible to add the book to the cart and finalize the purchase.
The detail pages for the two books in the example are of course different in terms of contained text, but from a functional testing view-point they are conceptually the same, as both are an instance  of the ``Show book details'' functionality.
Nevertheless, a ``naive'' crawler would assign them to different states, with negative consequences on subsequent model-based testing activities. For instance, test suites generated from these models with many near-duplicate states can be noticeably worse in terms of size and, running time \cite{yandrapally2020near}.

Despite being crucial for effective model inference activities, few research efforts have been directed towards detecting and discarding such near-duplicate web pages during the crawling process for E2E testing purposes. 
A first study in this direction was recently presented at ICSE 2020 \cite{yandrapally2020near}. In that study, 10 widely used similarity measures for web pages (e.g. Simhash \cite{charikar2002similarity}, which is used by Google during its indexing process) taken from different domains are applied and compared in the context of near-duplicate detection for web application model inference. The study shows that all of the techniques exhibited limitations and highlighted that there is a need for further research in devising novel approaches geared specifically towards model inference.

To address this issue, in our ongoing research we are investigating novel similarity measures for web pages, specifically designed for supporting model inference for web applications.
In particular, as web pages can be naturally represented using their tree-structured Document Object Model (DOM), we leverage Tree Kernel (TK) functions, a class of kernel functions largely investigated in the Natural Language Processing domain to evaluate similarity between tree-structured objects \cite{moschitti2006making}. 
We envision that TKs, thanks to their flexibility and customizability in the definition of what features to weight more for computing the similarity, might be effective tools to capture different types of near-duplicate web pages, improving the overall detection performance.

To understand the potentialities of our proposal, we conducted some preliminary experiments using a freely-available massive dataset of about 100k web page pairs, in which each pair was manually labelled as distinct, near-duplicate, or clone \cite{dataset-yandrapally}. We compared the classification performance of the proposed approach with the one achieved by the similarity measures investigated in \cite{yandrapally2020near}.
Preliminary results show that our TK-based approach outperforms all other techniques in the near-duplicate detection classification task.

The remainder of this paper is organized as follows. In Section \ref{sec:related} we survey related works on near-duplicate detection of web pages. In Section \ref{sec:tk-ndd} we introduce the TK-based near duplicate detection approach we propose, while in Section \ref{sec:empirical-evaluation} we describe the preliminary empirical evaluation we conducted.
In Section \ref{sec:results} we present the emerging results we obtained, and in Section \ref{sec:future} we provide some concluding remarks and a road-map detailing future research efforts.
\section{Related Works}\label{sec:related}
Many techniques from different domains have been defined, in different contexts, to the near-duplicate detection of web pages. 
For instance, the problem of detecting duplicate and near-duplicate web pages arises naturally in the web indexing process of search engines.
In this field, the concept of duplication and near-duplication is mainly related to the content of the web page, and hence Information Retrieval techniques have been found to be quite effective \cite{henzinger2006finding}.
Since performance is a crucial issue in this domain, due to the amount of involved data, content hashing techniques have been widely adopted thanks to their design simplicity and speed of comparison. Notable examples include the shingling algorithm presented by \citeauthor{broder1997syntactic} in \cite{broder1997syntactic}, and the Simhash algorithm \cite{charikar2002similarity}, which is also used by Google in its web page indexing process \cite{manku2007detecting}.
In \cite{henzinger2006finding}, \citeauthor{henzinger2006finding} carried out a large scale evaluation of these algorithms on a set of 1.6B distinct web pages, showing that both achieve high precision in detecting near-duplicate web page pairs across different websites, while performing significantly worse in detecting near-duplicated within the same website.

Detecting near-duplicate pages is also a challenge for automatic phishing detection. In this context, malicious websites are often designed to look as similar as possible to the original website they try to impersonate, while maintaining an entirely different HTML structure to avoid detection. Hence, techniques from the Computer Vision domain have often been applied to screen captures of the web pages with good results \cite{afroz2011phishzoo, wenyin2005detection}.

Among those visual-based techniques, the most fine grained approaches focus on  individual pixels composing the image. Examples of such techniques are \textit{color-histogram} \cite{swain1992indexing} and \textit{Perceptual Diff} (PDiff) \cite{yee2001spatiotemporal}, which have also been successfully applied in a previous web testing work for detecting cross-browser incompatibilities \cite{mahajan2014finding}. 
Some visual approaches operate at a coarser-grained scale, aiming at quantifying structural similarity or at extracting features from images. 
Structural similarity-based techniques leverage the intuition that images (and in particular screen captures of web applications) are typically highly structured, and their pixels, especially when they are spatially close, exhibit strong dependencies that convey important information about the structure of the represented objects. Similarity measures such as \textit{Structural Similarity Index} (SSIM) \cite{wang2004image}, which has been successfully applied in the detection of phishing websites \cite{chen2010detecting}, take into account these spatial correlations.

Other visual techniques are based on image hashing, aiming at computing identical or nearly-identical digests for similar images, e.g. the screen captures corresponding to near-duplicate web pages \cite{gangwar2018phishfingerprint}. Examples of image hashing algorithms include block-mean hash \cite{yang2006block} and perceptual hash (pHash) \cite{zauner2010implementation}.

Less work, however, has been directed towards the detection of near duplicate web pages \textit{within} the same web application and with the specific goal of supporting automated model inference.
Notable examples are the works presented in \cite{fard2013feedback,stocco2016clustering}, in which the authors apply the \textit{Robust Tree Edit Distance} (RTED) \cite{pawlik2015efficient} metric to respectively detect duplicate and near-duplicate web pages during dynamic crawling of web applications, and to reduce the size of the inferred models by clustering duplicate and near-duplicate states. 
More recently, in \cite{yandrapally2020near}, \citeauthor{yandrapally2020near} presented a comparative study in which 10 different near-duplicate detection techniques from different domains (including the aforementioned simhash, PDiff, color-histogram, SSIM, pHash, RTED) are applied and evaluated in the context of model inference. 
That study showed that none of the considered algorithms borrowed from the domains of information retrieval and computer vision ``\textit{is able to accurately detect all functional near-duplicates within apps}'', and ``\textit{underlined the need for further research in devising techniques geared specifically toward web test models}''.

\section{Near-duplicate Detection with Tree Kernels}\label{sec:tk-ndd}

As done in the work of \citeauthor{yandrapally2020near} \cite{yandrapally2020near}, we frame the near-duplicate detection problem as a multiclass classification problem. In particular, given a pair of web pages from the same web applications, the goal is to classify it into one of the following distinct categories:
\begin{itemize}
    \item \textbf{Clone}, if there is no semantic, functional or perceptible difference between two  web pages.
    \item \textbf{Distinct}, if there is any semantic or functional difference between the two pages.
    \item \textbf{Near-duplicate}, if there are noticeable differences, but the overall functionality being exposed is the same.
\end{itemize}
In this section, we start by giving some preliminary notions on Tree Kernel functions in \ref{sec:tk}, and then we introduce the Tree Kernel-based near-duplicate detection approach we are investigating in \ref{sec:proposed-approach}.

\subsection{Tree Kernel functions}\label{sec:tk}
\hyphenation{struc-tured}
Tree Kernel (TK) functions are a particular family of kernel functions which specifically evaluate similarity between two tree- structured objects. 
These functions have been extensively studied in Natural Language Processing \cite{moschitti2006making}, and have also been applied with promising results in the Software Engineering domain. In particular, TKs have been applied on Abstract Syntax Tree representations of source code for clone detection \cite{corazza2010tree}, and their usage is also being investigated for test case prioritization tasks \cite{altiero2020inspecting}.
More recently, \cite{ishikawa2020machine,shin2021learning} presented an effective approach to fake website detection, which leveraged TK functions. 

To compute the similarity between two trees $T_1$ and $T_2$, TK functions consider, for each tree, a set of \emph{tree fragments}. A tree fragment is a subset of nodes and edges of the original tree. 
Then, the similarity between the tree fragments of the two trees is evaluated, and the overall similarity of the two trees is computed by aggregating, in some meaningful way, the similarities of the single fragments.
Depending on how the set of fragments to consider is defined, 
it is possible to characterize different classes of tree kernel functions.
%
Widely-used classes include \cite{moschitti2006efficient}:
\begin{itemize}
    \item \textit{Subtree Kernels}, which consider only proper subtrees of the original trees, i.e., a node and all of its descendants, as fragments.
    \item \textit{Subset Tree Kernels}, which consider as fragments a more general structure than the one considered by subtree kernels, relaxing the constraint of taking all descendant of a given node and thus allowing for incomplete subtrees, limited at any arbitrary depth.
    \item \textit{Partial Tree Kernels}, which consider an even more general notion of fragment, in which the constraint of taking either all children of a tree node or none at all is relaxed. In this case, it is possible to include only some of the children of a node in a fragment. 
\end{itemize}

\subsection{The proposed approach}\label{sec:proposed-approach}

\begin{figure*}
    \centering
    \includegraphics[width=.7\linewidth]{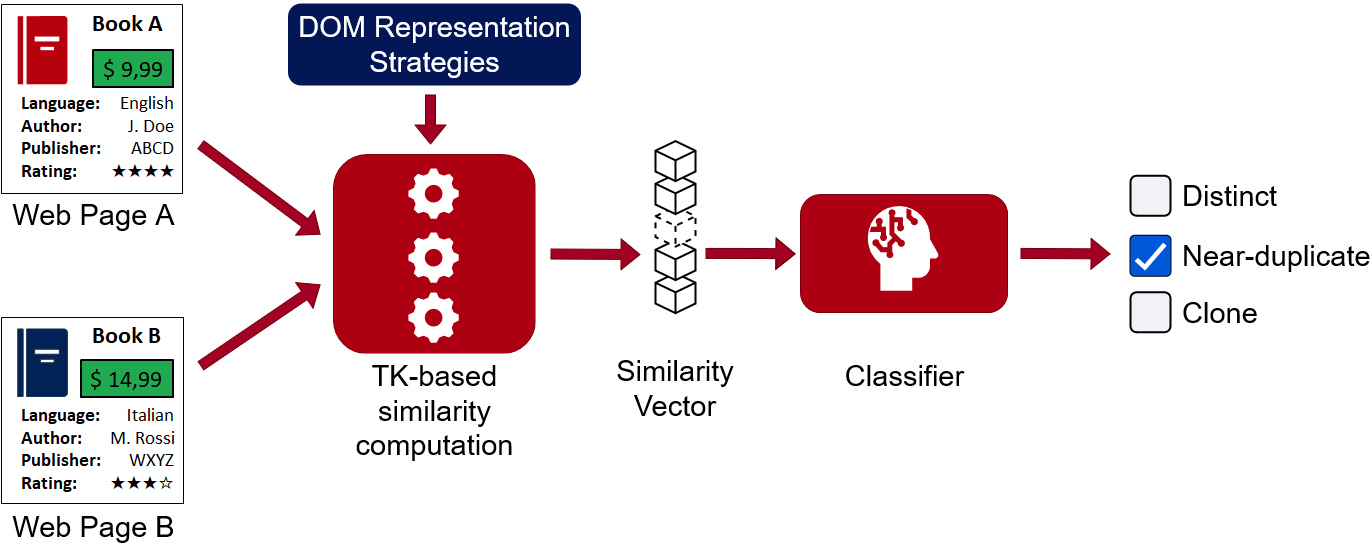}
    \caption{Overview of the proposed TK-based approach}
    \label{fig:approach}
\end{figure*}

\begin{figure}
    \centering
    \includegraphics[width=\linewidth]{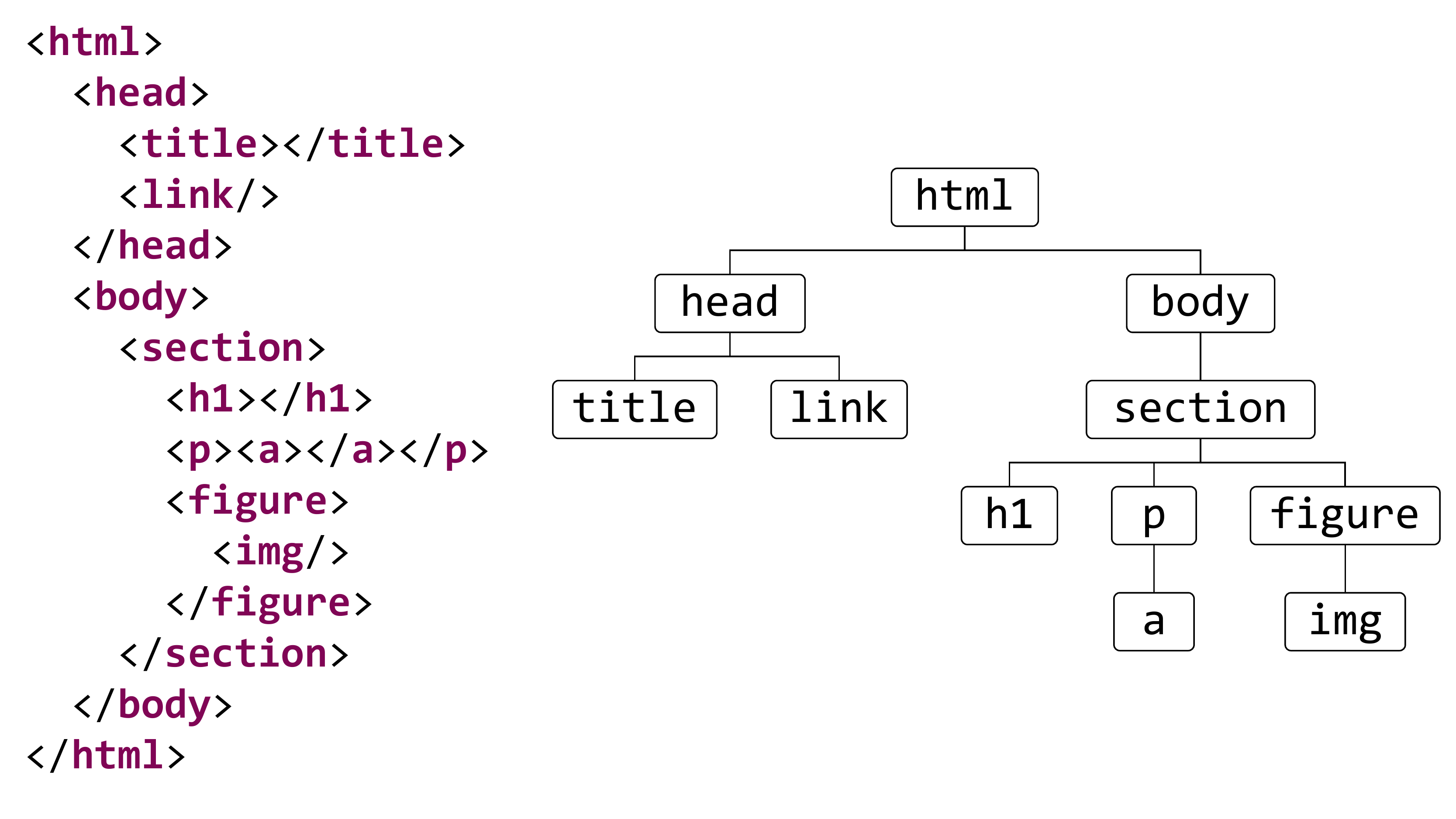}
    \caption{An HTML document and its DOM representation}
    \label{fig:html-dom}
\end{figure}

We envision that Tree Kernel (TK) functions might be an effective tool to measure the similarity of two web pages which, as shown in Figure \ref{fig:html-dom}, can be naturally modelled using their tree-structured Document Object Model (DOM) representation.
Since some preliminary experiments highlighted that none of the most common TK functions was able to detect all kinds of near-duplicates effectively, we devised a solution combining the similarity score computed by three different TK functions, namely a subtree kernel, a subset-tree kernel and a partial tree kernel.

Moreover, to investigate how different portions of the DOM tree impact similarity computation and near-duplicate detection, and to make our approach more general and customizable, we also introduce the concept of DOM representation functions. 
Intuitively, these functions represent a pre-processing step in which the DOM of a web page can be transformed according to some meaningful strategy.
Currently, we are considering three basic DOM representation strategies, as detailed in Table \ref{tab:dom-representations}.
\begin{table}
\caption{Considered DOM representation strategies}
    \label{tab:dom-representations}
    \small
    \centering
    \begin{tabular}{p{1.75cm}p{6cm}}
        \toprule
        \textbf{Strategy} & \textbf{Description} \\
        \midrule
         As-is & This representation strategy leaves the DOM unchanged; \\
         Only body & This representation strategy considers only the DOM subtree rooted in the \texttt{body} element of the web page.\\
         {Only body with no scripts} & This representation strategy is the same as the only body one, but also removes \texttt{script} elements along with their subtrees.\\
         \bottomrule
    \end{tabular}
\end{table}

From the pairwise combination of the three considered TK functions and the three DOM representation strategies, the similarity vector feeding the classifier has nine components for each pair of web pages.
Leveraging these similarity vectors and existing open datasets with annotated web page pairs, we use supervised learning approaches to train an \textit{ad-hoc} classifier. The proposed approach is summarized in Figure \ref{fig:approach}.

\section{Preliminary Empirical Evaluation}\label{sec:empirical-evaluation}

The goal of our preliminary experiment is to evaluate the effectiveness of the proposed TK-based approach in correctly detecting near-duplicate web pages, w.r.t. other state-of-the-art techniques.
To this end, we formulated and investigated the following research question:
\begin{description}
\item[RQ.] As compared to state-of-the-art approaches (i.e., the 10 techniques investigated in \cite{yandrapally2020near}), does the TK-based approach we propose allow achieving better near-duplicate detection performance?
\end{description}

\subsection{Employed Data}

We preliminarily evaluate the proposed approach using the same data and experimental procedure presented by \citeauthor{yandrapally2020near} in \cite{yandrapally2020near}. 
This way, our results can be directly compared with the state-of-the-art.
In particular, \cite{yandrapally2020near} compared 10 different near-duplicate detection techniques (which we use as a baseline) from the different domains of Computer Vision and Information Retrieval, and provided a large dataset of about 100k manually annotated same-website web page pairs, obtained by crawling both real-world websites and open source applications in a controlled environment.

The dataset they made available consists of three main parts detailed as follows: 
\begin{itemize}
    \item $\mathcal{SS}$ is a set of $\sim$97k annotated web page pairs extracted from 9 open source web applications in a controlled environment. The considered web applications were used in many previous works on web testing, and are briefly described in Table \ref{tab:apps}.\\
    \item $\mathcal{DS}$ is a set of $\sim$1k annotated same-website web page pairs extracted from about 1k real-world websites, randomly selected from Alexa's top 1 million URLs list.\\
    \item $\mathcal{TS}$ is a set of $\sim$500 additional annotated web page pairs extracted from the same websites as $\mathcal{DS}$.\\
\end{itemize}
\begin{table}
\caption{The considered open source web applications}
    \label{tab:apps}
    \small
    \centering
    \begin{tabular}{lp{5cm}l}
        \toprule
         \textbf{Web App} &
         \textbf{Description} \\
         \midrule
         Addressbook &  Simple address and phone book.\\
         PetClinic & Management of a veterinary clinic.\\
         Claroline & Collaborative e-learning platform.\\
         Dimeshift & Expense tracker.\\
         PageKit & Modular Content Management System.\\
         Phoenix & Project management.\\
         PPMA & Password Manager.\\
         MRBS & Meeting Room Booking System.\\
         MantisBT & Bug Tracker.\\
         \bottomrule
    \end{tabular}
    
\end{table}
For more details on the dataset and on the classification procedure the authors employed, we refer the interested reader to \cite{yandrapally2020near}.

\subsection{Experimental Procedure}

To evaluate the effectiveness of the TK-based classification approach we devised, we firstly extracted, for each web page pair in the dataset, the TK-based similarity vector we defined in Section \ref{sec:tk-ndd}. To do so, we leveraged the well-known open-source KeLP library, which features a state-of-the-art implementation of the tree kernel functions we employed \cite{filice2015kelp}.
Then, similarly to \cite{yandrapally2020near}, we used $\mathcal{DS}$ to train a SVM classifier, representing each web page pair with its computed similarity vector.
Finally, we evaluated classification performance on both $\mathcal{SS}$ and $\mathcal{TS}$, measuring for each of these datasets the macro-averaged $F_1$ classification score, as done in \cite{yandrapally2020near}.

The software component we implemented to extract the similarity vectors, as well as the R scripts we used to train and evaluate the SVM classifier, is open-source and available in the replication package at the public doi: \url{https://doi.org/10.6084/m9.figshare.14975178}. 

\subsection{Emerging results and discussion}\label{sec:results}

The results of this preliminary evaluation are reported in Table \ref{tab:results}, in which the first 10 rows show the best results obtained by the 10 state-of-the-art techniques in \cite{yandrapally2020near}, and the last one the results obtained by our TK-based approach.  
\begin{table}
\caption{Macro-averaged $F_1$ scores on $\mathcal{SS}$ and $\mathcal{TS}$}
    \label{tab:results}
    \small
    \centering
    \begin{tabular}{lrrr}
         \toprule
         \bfseries Technique & \bfseries $\mathcal{SS}$ & \bfseries $\mathcal{TS}$ & \bfseries Average\\
         \midrule
         PDiff & 0.53 & 0.67  & 0.60\\
         BlockHash & 0.54 & 0.62  & 0.58\\
         SSIM & 0.53 & 0.62  & 0.57\\
         Levenshtein & 0.48 & 0.59 & 0.54\\
         RTED & 0.50 & 0.57  & 0.54\\
         SIFT & 0.47 & 0.61  & 0.54\\
         pHASH & 0.40 & 0.63 & 0.52\\
         TLSH & 0.44 & 0.56 & 0.50\\
         Color-histogram & 0.37 & 0.52 & 0.44\\
         Simhash & 0.17 & 0.48 & 0.33\\
         \midrule
         TK-based SVM & 0.58 & 0.68 & 0.63\\
         \bottomrule
    \end{tabular}
    
\end{table}
These figures show that the proposed TK-based classification solution performs better than all the considered baseline approaches, achieving a 5\% improvement on $\mathcal{SS}$ and a 1\% improvement on $\mathcal{TS}$ w.r.t. \textit{Perceptual Diff} (PDiff), the best 
technique among those investigated in \cite{yandrapally2020near}.

It is worth noting that PDiff is a computationally expensive visual-based technique, and this can limit its application to model inference of large applications. 
Indeed, as reported in \cite{yandrapally2020near}, when using PDiff the model inference process could only explore 4 states per minute on average, as compared to faster, DOM-based approaches such as RTED, which could explore 25 states per minute.
The approach we propose is based on the DOM of the web pages, and thus is more efficient than PDiff.
When considering the best DOM-based techniques, namely RTED and the Levenshtein distance, our TK-based approach improves classification performance on both datasets by approximately 10\%.

\subsection{Threats to Validity}

\textit{External validity} threats concern the generalizability of the results. In this study, the employed dataset consists mainly of web page pairs from nine open-source web applications, which may not be representative of complex, real-world commercial applications. To mitigate this threat, the authors of the original dataset selected nine open source web applications from different domains, having different sizes, and implemented with different technologies. To further improve the generalizability of the results, in future works we plan to extend the original dataset by considering more web applications, including possibly commercial solutions.

A possible threat to \textit{internal validity}, concerning uncontrolled factors that may have affected the results, is represented by the manually created web page pair annotations. This threat is unavoidable, since there exists no automated method to compute the ideal classification of web pages. To minimize
this threat, the authors of the original dataset created, in isolation, a ground truth, and then established a discussion to reach an agreement \cite{yandrapally2020near}.
\section{Conclusions and Future Research}\label{sec:future}

Automated exploration techniques (a.k.a. crawling) are widely used to infer state-based models of web applications.
From a functional testing viewpoint, such inferred models should be as compact as possible, i.e., contain a minimal set of significantly different states, while maintaining completeness, adequately covering all the functionalities of the web application. 
In practice, however, models inferred automatically through state exploration are often affected by \textit{near-duplicate} states, i.e., states corresponding to replicas of the same functional web page differing only by minimal, insignificant changes.
This has a negative impact on the quality of the models and on the subsequent model-based testing tasks, adversely affecting, for example, size, running time, and achieved coverage of generated test suites.

In this paper, we introduced a novel approach to near-duplicate detection,  based on Tree Kernel functions, a class of kernel functions largely used in the Natural Language Processing domain to measure the similarity of tree-structured objects.
To preliminarily assess the effectiveness of the proposed approach, we conducted an empirical evaluation based on an open dataset of approximately 100k annotated web page pairs, in which we compared the near-duplicate detection performance of our approach against 10 baseline techniques.
Preliminary results were promising and showed that the TK-based approach performs better than all the baselines.

In future works, we plan to further investigate the potential of Tree Kernels in near duplicate detection and model inference along several research directions.
Firstly, we plan to improve the classification performance. Along this direction, we aim at designing custom TK functions specifically geared towards detecting near-duplicate web pages, as we believe that considering peculiar structural properties of web pages could make TKs more effective.
Furthermore, we also plan on extending the current approach, including more components in our similarity vectors. To this end, for example, we intend to experiment with more refined DOM representation strategies and with different kinds of TK functions, such as \textit{Subpath Kernels}, which were also recently applied, although in a different context, to web pages \cite{shin2021learning}.
We aim at implementing the solutions emerging from these studies as open-source extensions of the well-known Crawljax  web crawler \cite{mesbah2008crawling}, that will be made freely available to Software Engineering researchers and practitioners.
As for the empirical assessment, we plan to use the same data and experimental procedure used by \citeauthor{yandrapally2020near} in \cite{yandrapally2020near}, which provides a valuable state-of-the-art benchmark both for near-duplicate classification performances and for the quality of the inferred models.
Moreover, we plan to investigate the effectiveness of TK-based near-duplicate detection also in fully-automated E2E testing \cite{dimartino2021comparing} of mobile applications, leveraging the tree-like layout structure of their GUI.

\bibliographystyle{ACM-Reference-Format}
\bibliography{bibliography}

\end{document}